\documentclass[aps,prb,twocolumn,superscriptaddress,showpacs]{revtex4} 
\usepackage{amsmath}
\usepackage{amsfonts}
\usepackage{graphicx}

\begin{document}
%\preprint{version 1.0}

\title[]{Non linear transport theory for negative-differential
resistance states of two dimensional
electron systems in strong magnetic fields.}

\author{A. Kunold}
\email[ Email:]{akb@correo.azc.uam.mx, kunold@insa-toulouse.fr} 
\affiliation{Universit\'e de Toulouse; INSA-CNRS-UPS, LPCNO,
135, Av. de Rangueil, 31077 Toulouse, France}
\affiliation{Departamento de Ciencias B\'asicas,
Universidad Aut\'onoma Metropolitana-Azcapotzalco,
Av. San Pablo 180,  M\'exico D. F. 02200, M\'exico}
\author{M. Torres}
\email[Email:]{torres@fisica.unam.mx} 
\affiliation{Instituto de F\'{\i}sica,
Universidad Nacional Aut\'onoma de M\'exico,
Apartado Postal 20-364,  M\'exico Distrito Federal 01000, M\'exico}

\date{\today}
\begin{abstract}

We present a model to describe the nonlinear response to a direct  dc current
applied to a two-dimensional electron system in a strong  magnetic field.
The model is based on the solution of the von Neumann equation incorporating
the exact dynamics of two-dimensional damped electrons in the
presence of arbitrarily strong  magnetic  and  dc electric fields,
while the effects of randomly distributed impurities are perturbatively added.
From the analysis of the differential resistivity and the longitudinal voltage
we observe the formation of negative differential resistivity states (NDRS)
that are the precursors of the zero differential resistivity states (ZDRS).
The theoretical predictions  correctly reproduce the
main  experimental features  provided that the inelastic scattering  rate  obey a
$T^2$ temperature dependence, consistent with electron-electron interaction effects.
\end{abstract}

\maketitle

\section{Introduction}
In the past few years the study of non-equilibrium magneto-transport in high mobility
two-dimensional electron systems (2DES) has received much attention
due to the experimental finding of intense oscillations of the magneto-resistivity
and zero resistance states (ZRS).
Microwave-induced resistance oscillations (MIRO) were discovered
\cite{zudov:201311,zudov:046807,mani:646,ye:2193}
in 2DES samples subjected to microwave irradiation and moderate
magnetic fields. For the MIRO the photoresistance is a function of the
ratio $\epsilon^{ac}=\omega/\omega_c$ where $\omega$ and $\omega_c$
are microwave and cyclotron frequencies.
This outstanding discovery triggered a great amount of theoretical work
\cite{ryzhii:2078,ryzhii:165406,shi:086801,dorozhkin:577,durst:086803,
lei:226805,vavilov:035303,torres:115313,inarrea:073311,dmitriev:226802,
dmitriev:115316,dmitriev:165305,robinson:036804}.
Our current understanding
of this phenomenon rests upon models that predict the existence
of negative-resistance states (NRS) yielding an instability that
rapidly drive the system into a ZRS \cite{andreev:056803}. Two distinct mechanisms
for the generation of NRS are known, one is based in the
microwave-induced impurity scattering
\cite{ryzhii:2078,durst:086803,lei:226805,shi:086801,vavilov:035303,
torres:115313, inarrea:073311}, while the second
is linked to inelastic processes leading to  a non-trivial distribution function 
 \cite{dorozhkin:577,dmitriev:226802,dmitriev:115316,robinson:036804}.

An analogous effect,  Hall field-induced resistance oscillations (HIRO)
has been observed in high mobility samples in response to a dc-current
excitation \cite{yang:076801,zhang:081305,zhang:106804}.
Although MIRO and HIRO are basically different
phenomena both rely on the commensurability of the cyclotron frequency with a
characteristic parameter; in both cases oscillations are periodic in $1/B$.
In HIRO  the oscillation peaks, observed in differential resistance,
appear at integer values of the dimensionless parameter
$\epsilon^{dc} = \omega_H /\omega$.
Here, $\hbar \omega_H \approx  e E_H (2 R_C)$ is the energy associated
with the Hall voltage drop across the cyclotron diameter; $E_H$ is the
Hall field  and $R_C$ the cyclotron radius of the electron at the Fermi
level.
It has been found that there are two main contributions
to the HIRO: the {\it inelastic} one is related to the formation of a
non-equilibrium distribution function component that oscillates
as a function of the energy\cite{vavilov:115331}
and the {\it elastic} contribution is related to electron
transitions between different LLs due to impurity scattering\cite{lei:132119}.
The first one was shown to be dominant at relatively weak electric fields,
and the latter prevails in the strong-field regime.

More recently it has been demonstrated that the effects of a direct dc current
on electron transport  can be quite dramatic leading to zero
differential resistance states (ZDRS)\cite{bykov:116801,romero:153311}.
As compared with the HIRO conditions, the ZDRS are observed
under dc bias at  higher magnetic fields ($0.5-1.0 \, T$) and lower
mobilities ($70-85 \, m^2/V s$). At low temperature and above a threshold bias current
the differential resistivity vanishes and the
longitudinal dc voltage becomes constant. Positive values for the differential
resistance are recovered at higher bias as the longitudinal dc voltage  slope  becomes 
positive. Bykov et al. analyzed the results following  an approach similar to that
of Andreev et al. \cite{andreev:056803};
the presence of the ZDRS is attributed to the formation of negative differential
resistance states (NDRS) that yields an instability that drives
the system into a ZDRS. Similar results where obtained by Chen et al. \cite{chen:075308}

In this paper we  present a model to explain the formation of NDRS.
According to our formalism both the effects of elastic impurity scattering
as well as those related to inelastic processes play an important role.
The model is based on the solution of the von Neumann equation for 2D damped
electrons, subjected to arbitrarily strong magnetic and dc electric fields,
in addition to the weak effects of randomly distributed impurities.
This procedures yields a Kubo formula that includes the non-linear response
with respect to the dc electric field.
Considering a current controlled scheme, we obtain a set of nonlinear
self-consistent relations that allow us to determine the
longitudinal and Hall electric fields in terms of the imposed external current.
It is shown that in order to correctly reproduce the
main experimental results the inelastic scattering rate must obey a
$T^2$ temperature dependence, consistent with electron-electron
Coulomb interaction as the dominant inelastic process.

%FIGURE 1
%\begin{figure} [hbt]
\begin{figure}
\includegraphics[width=8 cm]{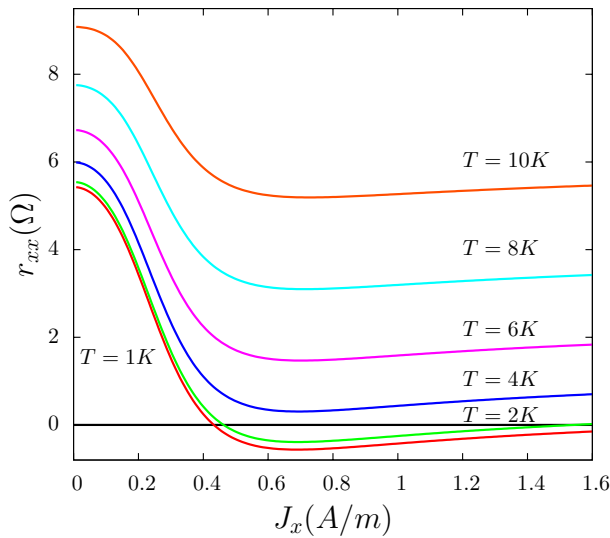}
\caption{Differential resistance $r_{xx}$ as a function of the
dc bias $J_x$ for $B=0.784 T$ temperatures from $T=1 K$ to $T=10 K$.}
\label{figure1}
\end{figure}

%FIGURE 2
%\begin{figure} [hbt]
\begin{figure}
\includegraphics[width=8 cm]{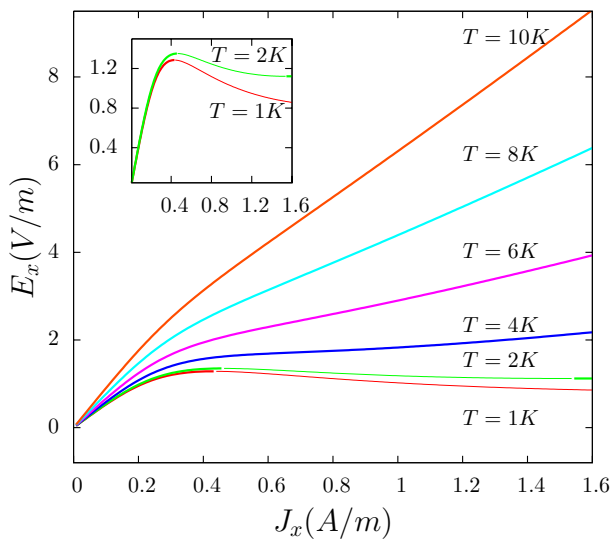}
\caption{Electric field $E_{x}$ as a function of the
dc bias $J_x$ for $B=0.784 T$ and for fixed temperatures ranging
from $T=1 K$ to $T=10 K$.}
\label{figure2}
\end{figure}

\section{Model}

We start with the Hamiltonian for an electron in the
effective mass approximation in two dimensions subject to a uniform
perpendicular magnetic field $\boldsymbol B=\left(0,0,B\right)$, an in-plane electric field
$\boldsymbol E=\left(E_x,E_y,0\right)$, and the  impurity scattering 
potential $V$. Hence the dynamics is governed by the total Hamiltonian 
$H = H_e + V$, with
\begin{equation}\label{ham0}
H_e=H_0
+e\boldsymbol{E}\cdot \boldsymbol x \, , 
\end{equation}
here $H_0=\boldsymbol{\Pi}^2/2m$, $m$ is the effective mass of the electron, $e$
is the electron's charge,
$\boldsymbol{\Pi}=\boldsymbol{p}+e\boldsymbol{A}$ is the velocity operator and the
vector potential  in the symmetric gauge is given as $\boldsymbol{A}=\left(-By,Bx\right)/2$.
The impurity scattering potential is expressed in terms of its Fourier
components
\begin{equation}\label{imppot}
V\left(\boldsymbol{r}\right)=
%\exp\left(-\eta\vert t\vert\right)
{\rm e}^{-\eta\left\vert t\right\vert}
\sum_i^{N_i}\int\frac{d^2q}{\left(2\pi\right)^2}
V\left(q\right)
\exp\left[i\boldsymbol{q}\cdot\left(\boldsymbol{r}
-\boldsymbol{r}_i\right)\right] \, ,
\end{equation}
where $\boldsymbol{r}_i$ is  the position of the $i$th impurity and
$N_i$ is the number of impurities.  The explicit form of
$V\left(q\right)$ depends on the nature of the scatterers\cite{torres:115313},
for simplicity we assume short-range uncorrelated scatterers. The factor
$\exp\left(-\eta\vert t\vert\right)$ takes care of the adiabatic switching of the
impurity potential at the initial time $t_0\to -\infty$.

The motion of a planar electron in  magnetic and electric fields
can be decomposed into the  guiding center coordinates
$\boldsymbol{Q}$ and  the relative coordinates
$\boldsymbol{R}=\left(-\Pi_y,\Pi_x\right)/e B$,
such that the position of the electron is given by
$\boldsymbol{r}=\boldsymbol{Q}+\boldsymbol{R}$.
The guiding center coordinates is written as
$\boldsymbol{Q}=\left({\cal Q}_x,{\cal Q}_y\right)/eB$,
in the symmetric gauge
$\left({\cal Q}_x,{\cal Q}_y\right)=\left(p_x+eBy/2,p_y-eBx/2\right)$.
The commutation relations for velocity and guiding center operators are
$\left[\Pi_x,\Pi_y\right]=\left[{\cal Q}_x,{\cal Q}_y\right]=-i\hbar eB$, with
all the other commutators being zero.

Our aim now is to compute  the  electric current density.
In order to calculate the expectation value  of
the current density we need the time-dependent  matrix $\rho(t)$ which obeys the
von Neumman's equation  $i\hbar\partial \rho/\partial t=\left[H,\rho\right]$.
We assume that in the absence of the impurity
potential the density matrix reduces to the equilibrium density matrix given
by $\rho_0 = f(H_0)$, with $f(E)$ given by the Fermi
distribution function.
In order to solve the von Neumman's equation we apply three unitary transformations:
the first two transformations exactly take into account the effects
of the electric and magnetic fields, whereas the third transformation incorporates
the  impurity scattering effects  to second order in time dependent
perturbation theory.
First we consider the unitary transformation
\begin{equation}\label{trans1}
{\cal W}\left(t\right)=e^{\frac{i}{\hbar}\int {\cal L}dt}
e^{-i\frac{v_x \Pi_y}{\hbar\omega_c}}e^{i\frac{v_y\Pi_x}{\hbar\omega_c}}
e^{i\frac{X {\cal Q}_x}{\hbar}}e^{i\frac{Y {\cal Q}_y}{\hbar}}
\end{equation}
where $v_x\left(t\right)$, $v_y\left(t\right)$, $X\left(t\right)$ and $Y\left(t\right)$ are
solutions of the dynamical equations
\begin{align}
& \dot{v}_x+\frac{1}{\tau_i}v_x+\omega_cv_y+\frac{e}{m}E_x=0,
& \dot{X}-\frac{Ey}{B}=0,\\
& \dot{v}_y+\frac{1}{\tau_i}v_y-\omega_cv_x+\frac{e}{m}E_y=0,
& \dot{Y}+\frac{Ex}{B}=0.
\end{align}
Except for the damping terms, these equations 
follow from the variation of the   classical Lagrangian
${\cal L}  $\cite{torres:115313}. The variables
$v_{x}$ and $v_{y}$ correspond to the  electron velocity components and $X$
and $Y$ are the coordinates that follow the drift of the electron's orbit.
In order to incorporate  dissipative effect  we added
the damping  term  $\boldsymbol{v}/\tau_i $  the  dynamical equations. This 
procedure yields a simple scheme to incorporate dissipation to the quantum system. 
Recent magnetoresistance experiments\cite{hatke:066804,hatke:161308}
and theory \cite{vavilov:115331} suggest, that  in 2DES,  electron-electron interaction 
provide an important contribution to the inelastic scattering rate, giving rise to 
$1/\tau_i \propto T^2$ temperature dependance.
Consequently, in what follows we shall assume that the inelastic scattering rate is given
by $ 1/ \tau_i\approx (k_BT)^2 / \hbar E_F$
\cite{chaplik:997,giuliani:4421,hatke:066804,hatke:161308}, where $E_F$ is the Fermi energy.

The transformation (\ref{trans1}) renders von Neumann equation
into the following form
\begin{equation}\label{vonneumann1}
i\hbar\frac{\partial \left({\cal W}\rho{\cal W}^{\dag}\right)}
{\partial t}\nonumber\\
=\left[H_0+V\left(t\right),{\cal W}\rho{\cal W}^{\dag}\right].
\end{equation}
The electric field term is conveniently removed from the Hamiltonian to
produce a time-dependent impurity potential
\begin{equation}
V\left(t\right)=V\left(x+X\left(t\right)+\frac{v_y\left(t\right)}{\omega_c},
y+Y\left(t\right)-\frac{v_x\left(t\right)}{\omega_c}\right).
\end{equation}
We proceed to switch to the interaction
picture through the unitary operator ${\cal U}_0=\exp\left(iH_0t/\hbar\right)$ and solve the
remaining equation up to second order in time dependent perturbation theory
obtaining yet another simplified version of von Neumann equation
\begin{equation}\label{vonneumann2}
i\hbar\frac{\partial}{\partial t} \left({\cal U}{\cal U}_0{\cal W}
\rho{\cal W}^{\dag}{\cal U}_0^{\dag}{\cal U}^{\dag}\right)=0,
\end{equation}
where the time evolution operator is  given by
\begin{align}
{\cal U}=&1-\frac{i}{\hbar}\int_{t_0}^tV_I\left(s_1\right)ds_1
\nonumber \\
&-\frac{1}{\hbar^2}\int_{t_0}^t\int_{t_0}^{s_1}
V_I\left(s_1\right)V_I\left(s_2\right)ds_1ds_2 \, , 
\end{align}
here $V_I\left(t\right)={\cal U}_0V\left(t\right){\cal U}_0^{\dag}$
is the impurity potential in the interaction picture.
The formal solution to (\ref{vonneumann2}) is given by
$\rho\left(t\right)={\cal W}^{\dag}{\cal U}_0^{\dag}{\cal U}^{\dag}
\rho\left(t_0\right){\cal U}{\cal U}_0{\cal W}$
where $\rho\left(t_0\right) = \rho_0 = f(H_0)$ is the equilibrium density
matrix at the initial time $t_0 \to - \infty $.

The density current is proportional to the thermal and time average
of the velocity operator 
\begin{eqnarray}
\boldsymbol{J}=\frac{e}{S}\int_{-\infty}^{\infty}dt{\rm Tr}\left[\rho\left(t\right) \boldsymbol{\Pi}\right] , 
\end{eqnarray}
where $S$ is the surface of the sample, and the limit $S \to \infty $ is understood.
By performing a cyclic permutation in the trace we obtain
\begin{equation}
\boldsymbol{J}=\frac{e}{S}{\rm Tr}\left[\rho\left(t_0\right)
{\cal U}{\cal U}_0{\cal W}\boldsymbol{\Pi}{\cal W}^{\dag}
{\cal U}_0^{\dag}{\cal U}^{\dag}\right]. 
\end{equation}
After lengthy calculations the components of the
density current is worked out as
\begin{multline}\label{denscurr}
J_i=\frac{ne^2\tau_i}{m}\frac{E_i-\omega_c\tau_i \epsilon_{ij}E_j}
{1+\omega_c^2\tau_i^2} \\
+\frac{e^2}{h}\sum_{\mu\mu^{\prime}}
\int d^2q
\left(f_{\mu}-f_{\mu^{\prime}}\right)
G^i_{\mu\mu^{\prime}}
\left(q\right)
\end{multline}
where $i,j=x,y$ and $\epsilon_{i,j}$ is the antisymmetric tensor
($\epsilon_{12}=-\epsilon_{21}=1$ and $\epsilon_{11}=\epsilon_{22}=0$),
\begin{multline}\label{dosexy}
G^i_{\mu\mu^{\prime}}=\frac{N_iB\left\vert V\left(q\right)
\right\vert^2}{Sm\hbar}\left\vert D_{\mu\mu^{\prime}}
\left(z_q\right)\right\vert^2\\
\frac{q_i \Delta_{\mu\mu^{\prime}}
+2\left\vert\epsilon_{ij}\right\vert q_j\omega_c\eta}
{\Delta^2_{\mu\mu^{\prime}}+4\omega_c^2\eta^2}
\end{multline}
and $\Delta_{\mu\mu^{\prime}}
=\left[\omega_q+\omega_c\left(\mu-\mu^{\prime}\right)\right]^2
-\omega_c^2+\eta^2$, $\omega_q=\omega_x E_x+\omega_y E_y$,
$\omega_x=-\tau_i\omega_c( q_x+q_y\tau_i \omega_c)/B(1+\tau_i^2\omega_c^2)$,
$\omega_y= \tau_i\omega_c(-q_y+q_x\tau_i \omega_c)/B(1+\tau_i^2\omega_c^2)$ and
$f_\mu=f\left(\hbar\omega_c\left(\mu+1/2\right)\right)$.
The matrix elements $D_{\mu,\nu}$ are given by
\begin{multline}\label{matD}
D_{\mu\mu^{\prime}} \left(z_q\right) =
\exp\left(-\frac{\left\vert z_q\right\vert^2}{2}\right)\\
\times
\left\{\begin{array}{ll}
z_q^{\mu-\mu^{\prime}}\sqrt{\frac{\mu^{\prime}!}{\mu!}}
L_{\mu^{\prime}}^{\mu-\mu^{\prime}}\left(\left\vert z_q \right\vert^2\right),&
\mu\ge\mu^{\prime},\\
\left(-{z_q}^*\right)^{\mu^{\prime}-\mu}
\sqrt{\frac{\mu!}{\mu^{\prime}!}}
L_{\mu^{\prime}}^{\mu^{\prime}-\mu}\left(\left\vert z_q\right\vert^2\right),&
\mu\le\mu^{\prime},\\
\end{array}\right.
\end{multline}
where  $z_q =(q_x-iq_y)/\sqrt{2}$ and $L_\nu^{\mu-\nu}$ denotes
the associated Laguerre polynomial.

Retaining a finite value of the switching  parameter  $\eta$ 
yields a density of states for the  Landau levels with the Lorentzian
form given in Eq. (\ref{dosexy}); it is distorted by the electric field through
the $\omega_q$ term. Henceforth we will consider $\eta=\Gamma \omega_c$.
The differential conductivity tensor is calculated  from
Eq. (\ref{denscurr}) as
$\sigma_{ij}=\partial J_i/\partial E_j$. Finally the  differential
resistivity tensor is obtained from the inverse of the conductivity:
that is $r_{ij}=\sigma^{-1}_{ij}$.

In the limit of small bias and small magnetic field the
expression for the density current reduces to
$J_x=ne^2\tau_iE_x\left(1-\alpha\right)/m$ where
\begin{equation}
\alpha=\frac{2\pi}{k_BT}
\frac{e^{-E_F/k_BT}}{\left(e^{E_F/k_BT}+1\right)^2}
\frac{\left\vert V\right\vert^2N_i m}{S\hbar \Gamma^2}.
\end{equation}
Hence the quantum scattering time and the inelastic scattering
time can be related by $\tau=\tau_i(1-\alpha)$ or similarly the elastic
scattering time is given by $\tau_e=\tau_i(1-\alpha)/\alpha$.
The factor $N_i\left\vert V\right\vert^2/S\Gamma^2$ present
in the expressions for the density current can be estimated
from the sample's mobility and the inelastic
scattering time.

In a current controled
scheme: the longitudinal density current is fixed to
a constant value $J_0$ while $J_y$ should vanish.
This leads to a set of two implicit equations for
the density current
\begin{eqnarray}\label{condi}
J_x\left(E_x,E_y\right)=J_0, && J_y\left(E_x,Ey\right)=0,
\end{eqnarray}
where the  explicit  form of the  functions $J_i$ is given  in Eq. (\ref{denscurr}).
To obtain the components of the electric field $E_x$ and $E_y$, we  start
assigning  initial  values  $E_x=E_{x_0}$ and $E_y=E_{y_0}$ that solve
these relations in the absence of impurities ($i.e.$   using only
the first term on the R.H.S. of Eq. (\ref{denscurr})), then  the accuracy of
the solution  is improved by a recursive application of Newton's method.

%FIGURE 3
%\begin{figure} [hbt]
\begin{figure}
\includegraphics[width=8 cm]{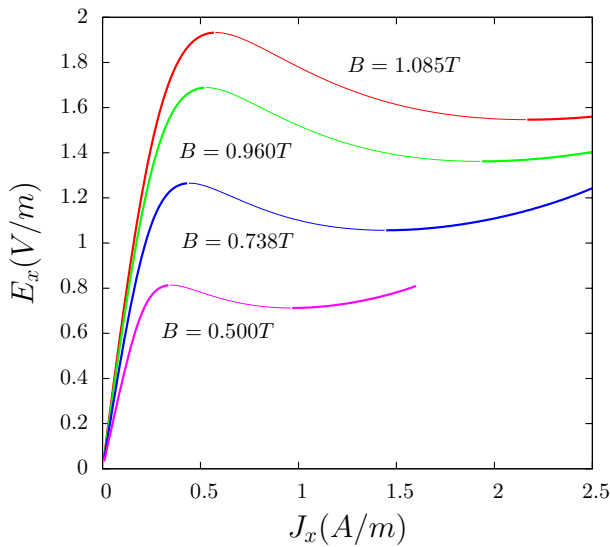}
\caption{Electric field $E_{x}$ as a function of the
dc bias $J_x$ for $T=2 K$ and for fixed magnetic fields
ranging from $B=0.5 T$ to $B=1.085 T$. The thin lines
indicate that $r_{xx}<0$.}
\label{figure3}
\end{figure}

%FIGURE 4
%\begin{figure} [hbt]
\begin{figure}
\includegraphics[width=8 cm]{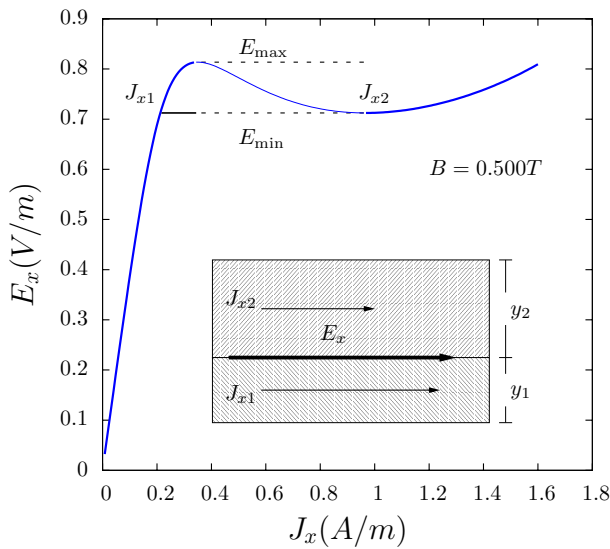}
\caption{Electric field $E_{x}$ as a function of the
dc bias $J_x$ for $T=2 K$ and for magnetic field
$B=0.5 T$. The thin lines
indicate differential resistivity $r_{xx}<0$. The inset shows a
possible non uniform configuration for the density current. }
\label{figure4}
\end{figure}

\section{Results}
 Fig. \ref{figure1} shows  the differential resistivity
$r_{xx} = \partial E_x / \partial J_x $ as a function of the longitudinal
dc density current $J_x$ for  a magnetic field  $B=0.784 T$ and various values
of the temperature. We use a sample mobility  $\mu=100 m^2 V/s$, electron density
$n=8.2\times 10^{15}m^{-2}$ and a broadening parameter  $\Gamma=0.04$.
As the value of the temperature is reduced the differential resistance  decrease approaching zero.
We can observe that at  low temperature ($T<2K$) and above a threshold bias current ($J_x>0.4 A/m$)
the differential resistivity becomes negative. Positive values for the differential
resistance are recovered at higher bias or higher temperatures.
The strong temperature dependence observed in this plots, consistent with the experiments,
is originated mainly on the  $T^2$ dependence of the inelastic scattering rate.

The electric field $E_x$ is plotted as a function of the longitudinal current
$J_x$ in Fig. \ref{figure2}. It is important to notice that $E_x$ differs from
the longitudinal voltage by a geometrical factor. DNRS are observed below $T=4K$
and above the  current threshold $J_x>0.4 A/m$ in the form of negative slope curves
(see inset of Fig. \ref{figure2}) in accordance
with the $r_{xx}$ negative values observed in Fig. \ref{figure1}.
According  to Bykov et al. \cite{bykov:116801}  the stability condition
is simply expressed as $r_{xx}\ge 0$. Thus the regions in Figs. \ref{figure1} and
\ref{figure2} that display a negative differential resistivity are unstable,
and they  should  rapidly evolve into ZDRS to insure stability.
Accordingly in Fig. \ref{figure1} we should  replace  the NDRS by $r_{xx}=0$ 
and maintain a constant slope in Fig. \ref{figure2} instead of the negative slope.
At higher values of $J_x$ the differential resistivity becomes
positive (Fig. \ref{figure1}) as well as the longitudinal voltage slope
as a result of an increase in the impurity scattering prevalent
at high electric fields. In this regime the large
electric field components, necessary to maintain the strong dc bias and $J_y=0$,
cause the impurity terms to strongly participate\cite{vavilov:115331}.

Fig. \ref{figure3}  display a series of  plots of  $E_x$ field as a function of the
longitudinal density current $J_x$ at $T=2K$  for various fixed values of the magnetic
field that correspond to Shubnikov-de Haas oscillations maxima.
The thin lines indicate negative values of $r_{xx}$ that violate
the stability condition. As the magnetic field increases the width of the electric
field plateaus  increase and the positive slope is recovered for higher onset
density currents.
An isolated plot of the longitudinal electric field $E_x$ as a function of the
dc current $J_x$ is shown in Fig. \ref{figure4}.
In the inset of Fig. \ref{figure4} we show a nonuniform distribution current
similar to the one proposed by Bykov et al.\cite{bykov:116801}.
With this configuration not only the stability condition $r_{xx}>0$ is fulfilled
but the electric field is uniform throughout the sample given that
$E_x=E_{\rm min}$ for $J_{x1}$ and $J_{x2}$.
The average current density
$J_x=(J_{x1}y_1+J_{x2}y_2)/(y_1+y_2)$ may be modulated by varying the sizes
 $y_1$ and $y_2$ of the different density current domains with
the restriction that $y_1+y_2=w$. Notice that more complicated schemes with more density
current modulations also fulfill this conditions.

\section{Conclusions}
We have presented a model for the nonlinear transport of a 2DES placed in
a strong  perpendicular magnetic field.
The model is based on the solution of the von Neumann equation for 2D damped electrons,
subjected  to arbitrarily  strong magnetic and  dc electric fields, in addition to the
weak effects of randomly distributed impurities. This procedures yields a Kubo formula
that includes the non-linear response with respect to  the dc electric field.
Considering a current controlled scheme, we obtain a set of nonlinear self-consistent
relations that allow us to determine the longitudinal and Hall electric fields
in  terms of the imposed external current.
NDRS are found in the low temperature ($T\le 2$) and moderate bias regime
$0.4 A/m<J_x<1.6 A/m $. In low dc bias (low electric field regime) the dominant
mechanism is the inelastic one. The longitudinal electric field (and voltage)
recover they positive slope in the high bias (high electric field regime).
It is shown that in order to correctly reproduce the main experimental results
the inelastic scattering rate must obey a $T^2$ temperature dependence,
consistent with electron-electron Coulomb interaction as the dominant inelastic process.

\acknowledgments
A. Kunold is receiving financial support from
``Estancias sab\'aticas al extranjero'' CONACyT and ``Acuerdo 02/06'' Rector\'{\i}a UAM-A.
A. Kunold wishes to thank INSA-Toulouse for his hospitality.

%\bibliography{kunold}

\end{document}